\documentstyle[prd,floats,eqsecnum,aps,axodraw]{revtex}

\newcommand{\SM}{SU(3) \otimes SU(2) \otimes U(1)}

\begin{document}

\preprint{ULB-TH-01/025}
\draft

\def\today{\space\number\day\space\ifcase\month\or January\or February\or
  March\or April\or May\or June\or July\or August\or September\or October\or
  November\or December\fi\space\number\year}
 
\twocolumn[\hsize\textwidth\columnwidth\hsize\csname@twocolumnfalse\endcsname  

\title{Anomalies and Fermion Content of Grand Unified Theories in 
Extra Dimensions}

\author{Nicolas Borghini,
Yves Gouverneur,
and Michel H.\ G.\ Tytgat}

\address{Service de Physique Th\'eorique, CP225, 
Universit\'e Libre de Bruxelles, B-1050 Brussels, Belgium}

\date{10 August 2001}
\maketitle

\begin{abstract}
The restrictions imposed by anomaly cancellation on the chiral fermion content 
of nonsupersymmetric gauge theories based on various groups are studied in 
spacetime dimension $D=6$, 8, and 10. 
In particular, we show that the only mathematically consistent chiral $SU(5)$ 
theory in $D=6$  contains three nonidentical generations.  
\end{abstract}

]

\narrowtext

\section{Introduction}
\label{s:intro}

Despite its numerous successes, the Standard Model of particle physics is far 
from being satisfactory. 
The fermion sector is particularly puzzling. 
Among other problems, one may wonder why there are so many 
different fermions, with apparently arbitrary quantum numbers under $\SM$, 
and why it is possible to divide them into three generations. 

The first question can be partially answered: the quantum numbers ensure the 
cancellation of all potentially dangerous chiral anomalies 
\cite{Weinberg:1996kr}. 
Historically, the latter were discovered \cite{ABJ} while most fermions of 
the Standard Model were already known experimentally: the absence of anomaly 
was more a way of checking the consistency {\em a posteriori\/} than a predictive 
tool. 
Nevertheless, the anomaly cancellation condition led to an alternative
prediction of the existence of the $c$ quark \cite{Bouchiat:1972iq}. 
Furthermore, it has been shown that, with some additional assumption, namely, 
that the fermions may only be $SU(2)$ [resp.\ $SU(3)$] singlets or doublets 
(resp.\ triplets), an anomaly-free fermion content with the minimal 
number of fields fits precisely within one generation of the Standard Model 
\cite{Geng:1989pr}. 
However, four-dimensional anomalies do not explain why there should be three 
generations in Nature. 

Various explanations for the existence of several generations have been 
proposed. 
In theories with extra dimensions, for instance,  the number of generations 
can be related, through the index theorem, to the topology of the compact 
manifold \cite{Witten:1981me} or to the winding of some field configuration 
(see e.g.\ \cite{Frere:2001dc}). 
In the Connes-Lott version of the Standard Model in noncommutative geometry, 
the existence of spontaneous chiral symmetry breaking requires the existence 
of more than one generation \cite{Schucker:1995xj}. 
Recently, it has been proposed that anomalies could actually yield a 
constraint on the number of generations, provided the cancellation of 
anomalies takes place in an $\SM$ theory that lives in six spacetime 
dimensions ($6D$)\cite{Dobrescu:2001ae}. 

In this paper, we shall further investigate the anomaly cancellation condition 
in arbitrary spacetime dimension, extending the discussion of 
\cite{Dobrescu:2001ae} to larger groups containing $\SM$. 
We shall only consider the case of even dimensions: in odd dimensions, there 
is no chirality, hence no chiral anomaly, and the closest equivalent, the 
parity anomaly, can be canceled by a Chern-Simons term in the action 
\cite{Redlich:1984kn}. 
Also, since anomalies yield no information on vector-like generations, we 
shall only derive constraints on the number $n_g$ of chiral generations. 

In order to make our paper self-contained, we first review in 
Sec.\ \ref{s:anom_types} the different types of chiral anomalies which will be 
relevant in the sequel. 
In Sec.\ \ref{s:constraints}, we impose the absence of anomaly in 
(nonsupersymmetric) gauge theories based on any of the groups $\SM$, $SU(5)$, 
$SO(10)$, and $E_6$, in dimensions $D=6$, 8, and 10, and deduce in each case 
the possible fermion contents. 
Among the various cases, the $SU(5)$-based theories are the most constrained: 
in particular, in six dimensions, only theories with $n_g=0$ mod 3 generations 
are anomaly-free.  
Let us emphasize rightaway that, because charge conjugation does not change 
chirality in $D=6$, this $SU(5)$ solution is not a trivial generalisation of 
the well-known $4D$ construction.  
In Sec.\ \ref{s:model}, we study the possible embedding of the $4D$ Standard 
Model in this $6D$ $SU(5)$ theory.  
We give some conclusions and prospects for future works  in 
Sec.\ \ref{s:conclusions}. 
Finally, some useful results and demonstrations are given in the Appendixes 
\ref{s:C,P,etal} and \ref{s:Tr(SO10)^5}.

\section{Short review of chiral anomalies}
\label{s:anom_types}

A symmetry is said to be anomalous if it exists at the classical level, 
but does not survive quantization. 
In some cases anomalies are welcome, as in the $\pi^0$ decay \cite{ABJ}. 
These harmless anomalies are always associated with global symmetries of 
the Lagrangian. 
In opposition, anomalies which affect local symmetries, in particular gauge 
symmetries, jeopardize the theory consistency.
Such anomalies spoil renormalizability; but even in the case of effective, 
{\it a priori} non-renormalizable theories, they destroy unitarity, leading to 
theories without predictive power \cite{'tHooft:1979bh}. 
Consistent models should therefore either contain none of these anomalies, 
or automatically cancel them \cite{Georgi:1972bb,Okubo:1977sc}.
Conversely, the cancellation condition gives useful constraints on the 
structure of a theory, and especially on its fermion content 
\cite{Gross:1972pv}, as we shall recall in Sec.\ \ref{s:constraints}.

We shall be interested in the so-called chiral anomalies, which involve 
chiral fermions in the presence of gauge fields and/or gravitons. 
They can be divided in two classes: local (Sec.\ \ref{s:local}) and global 
(or nonperturbative; see Sec.\ \ref{s:global}), according to 
whether they can be calculated perturbatively or not.

\subsection{Local anomalies}
\label{s:local}

\begin{figure}[ht!]
  \begin{center}
    \begin{picture}(180,80)(0,0)
      \ArrowLine(35,60)(21,36)
      \ArrowLine(21,36)(49,36)
      \ArrowLine(49,36)(35,60)
      \Photon(35,80)(35,60){3}{3}
      \Photon(0,25)(21,36){3}{3}
      \Photon(70,25)(49,36){3}{3}
      \Text(35,10)[b]{anomaly for $D=4$}
      \ArrowLine(138,62)(138,38)
      \ArrowLine(138,38)(162,38)
      \ArrowLine(162,38)(162,62)
      \ArrowLine(162,62)(138,62)
      \Photon(120,80)(138,62){3}{3}
      \Photon(120,20)(138,38){3}{3}
      \Photon(180,80)(162,62){3}{3}
      \Photon(180,20)(162,38){3}{3}
      \Text(150,10)[b]{anomaly for $D=6$}
    \end{picture}
    
    \begin{picture}(180,90)(0,0)
      \ArrowLine(35,68)(18,56)
      \ArrowLine(18,56)(24,35)
      \ArrowLine(24,35)(46,35)
      \ArrowLine(46,35)(52,56)
      \ArrowLine(52,56)(35,68)
      \Photon(35,90)(35,68){3}{3}
      \Photon(0,64)(18,56){3}{3}
      \Photon(12,19)(24,35){3}{3}
      \Photon(58,19)(46,35){3}{3}
      \Photon(70,64)(52,56){3}{3}
      \Text(35,8)[b]{anomaly for $D=8$}
      \ArrowLine(126,52)(135,36)
      \ArrowLine(135,36)(153,36)
      \ArrowLine(153,36)(162,52)
      \ArrowLine(162,52)(153,68)
      \ArrowLine(153,68)(135,68)
      \ArrowLine(135,68)(126,52)
      \Photon(108,52)(126,52){3}{3}
      \Photon(126,22)(135,36){3}{3}
      \Photon(162,22)(153,36){3}{3}
      \Photon(180,52)(162,52){3}{3}
      \Photon(162,82)(153,68){3}{3}
      \Photon(126,82)(135,68){3}{3}
      \Text(144,8)[b]{anomaly for $D=10$}
    \end{picture}
    \caption{Anomalous diagrams in $D=4$, 6, 8, and 10 dimensions.
      Each external leg stands for any of the gauge bosons of the theory, 
      while the fermions circulating in the internal lines can be in any 
      relevant representation of the gauge group.}
    \label{fig:diag4D-10D}
  \end{center}
\end{figure}
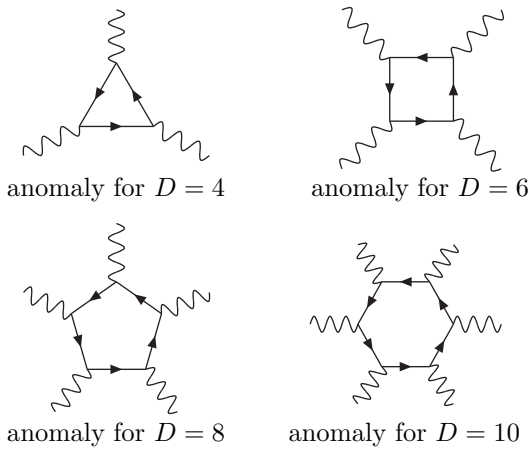

Local anomalies are related to infinitesimal gauge and/or coordinate 
transformations. 
They arise from a typical kind of Feynman diagram, which leads to a possible 
non-conservation of the gauge symmetry current or the energy-momentum tensor. 
The topology of these diagrams depends on the spacetime dimension $D$. 
In $D=4$, these are the well-known triangle diagrams \cite{ABJ}. 
In 6-, 8-, and 10-dimensional theories, the corresponding possibly anomalous 
diagrams are respectively the so-called box, pentagonal, and hexagonal 
diagrams \cite{Frampton:1983nr,Zumino:1984rz}, represented together with the 
triangle diagram in Fig.\ \ref{fig:diag4D-10D}. 

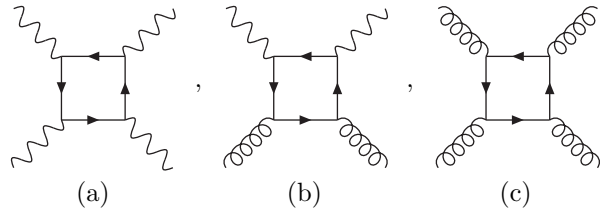
\begin{figure}[ht!]
  \begin{center}
    \begin{picture}(220,80)(0,0)
      \ArrowLine(18,62)(18,38)
      \ArrowLine(18,38)(42,38)
      \ArrowLine(42,38)(42,62)
      \ArrowLine(42,62)(18,62)
      \Photon(0,80)(18,62){3}{3}
      \Photon(0,20)(18,38){3}{3}
      \Photon(60,80)(42,62){3}{3}
      \Photon(60,20)(42,38){3}{3}
      \Text(30,10)[]{(a)}
      \Text(70,50)[]{$,$}
      \ArrowLine(98,62)(98,38)
      \ArrowLine(98,38)(122,38)
      \ArrowLine(122,38)(122,62)
      \ArrowLine(122,62)(98,62)
      \Photon(80,80)(98,62){3}{3}
      \Gluon(80,20)(98,38){3}{4}
      \Gluon(140,20)(122,38){-3}{4}
      \Photon(140,80)(122,62){3}{3}
      \Text(110,10)[]{(b)}
      \Text(150,50)[]{$,$}
      \ArrowLine(178,62)(178,38)
      \ArrowLine(178,38)(202,38)
      \ArrowLine(202,38)(202,62)
      \ArrowLine(202,62)(178,62)
      \Gluon(160,80)(178,62){3}{4}
      \Gluon(160,20)(178,38){3}{4}
      \Gluon(220,80)(202,62){-3}{4}
      \Gluon(220,20)(202,38){-3}{4}
      \Text(190,10)[]{(c)}
    \end{picture}
    \caption{Local [(a) pure gauge, (b) mixed, and (c) pure gravitational] 
      anomalies in 6 dimensions. Wavy external legs stand for gauge bosons, 
      while loopy legs stand for gravitons.}
    \label{Fig:6Danomalies}
  \end{center}
\end{figure}

One can distinguish three types of local anomalies, according to the nature 
of the external legs of the anomalous diagrams. 
Diagrams with only gauge bosons correspond to the pure gauge anomaly. 
On the other hand, when all external legs are gravitons, the diagram yields 
the pure gravitational anomaly \cite{Alvarez-Gaume:1984ig}. 
Finally, the mixed anomaly correspond to diagrams with both gauge bosons and 
gravitons \cite{Alvarez-Gaume:1984ig,Delbourgo:1972xb}. 
These various types are illustrated, in the case of $D=6$ dimensions, in 
Fig.\ \ref{Fig:6Danomalies}. 

\subsubsection{Pure gauge anomaly}

In the pure gauge case, the anomaly is proportional to a group factor, 
which multiplies a Feynman integral: 
\begin{equation}
\label{gaugeDD}
\sum_{L_D} {\rm STr} \left( T^a T^b \ldots T^{\frac{D}{2}+1} \right) - 
\sum_{R_D} {\rm STr} \left( T^a T^b \ldots T^{\frac{D}{2}+1} \right),
\end{equation}
where the notation ${\rm STr}$ means that the trace is performed over the 
symmetrized product of the gauge group generators $T^a$. 
This symmetrization is related to the Bose-Einstein statistics of the 
interaction fields. 
The sums run over all left- and right-handed (in the $D$-dimensional sense, 
see Appendix \ref{s:C,P,etal}) 
fermions of the theory belonging to the representation $T^a$ . 

The symmetrized traces of Eq.\ (\ref{gaugeDD}) can be expressed in terms of 
traces over products of the generators $t^a$ of the fundamental 
representation, and can sometimes be factorized. 
The first property reduces the number of traces which must be calculated, 
provided the coefficients relating the traces over arbitrary generators to 
the traces over the $t^a$ are known \cite{Frampton:1983nr,Okubo:1983sv}. 
The second property is due to the existence of basic (i.e., non-factorizable) 
traces, the number of which, and the number of generators they involve, being 
related to the rank and the Casimir operators of the group. 
A simple example is $SU(2)$, which is of rank 1, with the unique Casimir 
operator $(T)^2$; a trace involving more than two generators can be factorized: 
\begin{eqnarray}
  &  {\rm STr}\left(T^aT^bT^c\right) \propto  
  {\rm S}\left( {\rm Tr}(T^aT^b) \, {\rm Tr}\,T^c \right) = 0, & \\
  \label{SU2-4D}
  & {\rm STr}\left(T^aT^bT^ cT^d\right) \propto 
  {\rm S}\left({\rm Tr}(T^aT^b) \, {\rm Tr}(T^cT^d)\right). & 
  \label{SU2-6D}
\end{eqnarray}
In other words, the triangle diagram for $[SU(2)]^3$ (sometimes called cubic 
anomaly) vanishes for any $SU(2)$ representation of fermions. 
Equation (\ref{SU2-6D}) states that the quartic $SU(2)$ anomaly is 
factorizable.

The pure gauge anomaly (\ref{gaugeDD}) vanishes either if the group is 
``safe'' \cite{Georgi:1972bb,Okubo:1977sc}, as is the case for $SU(2)$ in four 
dimensions, or if the fermion content of the theory is properly chosen. 
Nevertheless, it is possible that part of the anomaly is zero thanks to the 
matter content, while the remaining part can be canceled by an additional 
tensor, through the Green-Schwarz mechanism \cite{Green:1984sg}, as will be 
discussed later.

\subsubsection{Pure gravitational anomaly}

The gravitational anomaly \cite{Alvarez-Gaume:1984ig} represents a breakdown 
of general covariance, or, equivalently, of the conservation of the 
energy-momentum tensor, due to parity-violating couplings between fermions and 
gravitons. 
In particular, chiral fermions obviously violate parity, and lead to such 
anomalies. 
A necessary and sufficient condition for the absence of local gravitational 
anomaly is therefore the identity of the numbers of left- and right-handed 
fermions:
\begin{equation}
\label{gravitational}
N_{L_D}-N_{R_D}=0. 
\end{equation}

As we recall in Appendix \ref{s:C}, in dimension $D=4k$ charge conjugation 
flips chirality, while it does not in $D=4k+2$. 
Therefore, a left-handed Weyl fermion field contains a left-handed particle 
and an antiparticle with opposite (resp.\ identical) chirality in $D=4k$ 
(resp.\ $D=4k+2$).
Such a field is, from the gravitational point of view, vector-like in dimension 
$4k$, while it is chiral in dimension $4k+2$. 
Thus, the local gravitational anomaly always vanishes if the spacetime 
dimension is $D=4k$. 

Note that a gravitational anomaly can always be canceled by the addition of 
the right number of gauge singlet chiral fields. 
This addition obviously does not affect the gauge and mixed anomalies. 
Therefore, this anomaly does not yield a very stringent constraint from the 
phenomenological point of view.

\subsubsection{Mixed anomaly; Green-Schwarz mechanism}

The mixed gauge-gravitational anomaly [Fig.\ \ref{Fig:6Danomalies}, diagram 
(b)] is proportional to the product of a gauge group factor and a 
gravitational term. 
The latter vanishes when the number of gravitons is odd 
\cite{Alvarez-Gaume:1984ig}. 

When the mixed anomaly does not vanish thanks to group properties or an 
appropriate fermion choice, it may still be canceled through the Green-Schwarz 
mechanism \cite{Green:1984sg}.
This mechanism relies on the existence, in dimension $D\geq 6$, of tensors 
which, with properly chosen couplings, can cancel anomalies proportional to 
the trace of the product of $k$ generators, with $2 \leq k \leq D/2-1$. 
The anomalies which can be canceled in this way, either mixed or pure gauge, 
are called reducible, and the others, irreducible. 
Since gauge and mixed anomalies can be factorized, the factorization may 
amount to dividing the anomaly into a reducible part, which can be canceled 
through the Green-Schwarz mechanism, and an irreducible part,\footnote{In 
fact, a {\em sufficient\/} condition for the existence of an irreducible 
anomaly for the gauge group $G$ is $\Pi_{D+1}(G)=Z$, where $\Pi_{D+1}(G)$ is 
the ($D+1$)-th homotopy group of $G$ \cite{Kiritsis:1986mf}.} which 
necessitates some appropriate fermion content.

\subsection{Global anomalies}
\label{s:global}

In addition to the local anomalies discussed previously, there are also 
nonperturbative anomalies, which cannot be obtained from a perturbative 
expansion, and will be called global in the following, although they are 
related to local symmetries. 
Two types of such anomalies can arise, related either to gauge invariance or 
to gravity. 

The global gauge anomaly \cite{Witten:1982fp} occurs when there exist gauge 
transformations which cannot be deduced continuously from the identity, in 
the presence of chiral fermions. 
In other terms, the anomaly arises when the $D$-th homotopy group of the gauge 
group $G$, $\Pi_D(G)$, is nontrivial. 
The anomaly then leads to mathematically inconsistent theories in which all 
physical observables are ill-defined. 

This anomaly vanishes only if the matter content of the theory is appropriate. 
More precisely, if $\Pi_{D}(G)\neq 0$, the cancellation of the anomaly 
constrains the numbers $N(p_{L_D})$ and $N(p_{R_D})$ of left- and right-handed 
$p$-uplets: 
\begin{equation}
\label{global}
\Pi_D(G)=Z_{n_D} \Rightarrow  
c_D \left[N(p_{L_D})-N(p_{R_D})\right]=0 \mbox{ mod } n_D,
\end{equation}
where $c_D$ is an integer whose value depends on the spacetime dimension $D$, 
the gauge group $G$, and the representation of $G$ the fermions belong to. 

In the case of the $SU(2)$ global anomaly in $D=4$ dimensions, 
$\Pi_{4}(SU(2))=Z_2 $,  so the anomaly cancellation condition reads, for the 
fundamental representation ($c_4=1$),  
$N(2_{L_4})-N(2_{R_4}) = 0 \mbox{ mod } 2$, 
where $N(2_{L_4})$ and $N(2_{R_4})$ are the numbers of left- and right-handed 
Weyl fermions which are doublets under $SU(2)$ \cite{Witten:1982fp}. 

Coordinate transformations which cannot be reached continuously from the 
identity give rise to possible global gravitational anomalies 
\cite{Alvarez-Gaume:1984ig}. 
In a $(4k+2)$-dimensional spacetime, these anomalies vanish when condition 
(\ref{gravitational}) holds: the cancellation of the local gravitational 
anomaly automatically ensures that the global one is zero. 
In $D=8k$ dimensions, the anomaly vanishes only if the number of (spin 
$\frac{1}{2}$) Weyl fermions coupled to gravity is even; otherwise, the 
theory is inconsistent. 
Note that in that case, there is no local gravitational anomaly. 
This is similar to the possibility of global gauge anomalies for $SU(2)$ in 
4 dimensions, while there is no corresponding local anomaly. 

Finally, there is another important feature of anomalies, which we shall 
encounter in the following, related to symmetry breaking. 
When a symmetry is spontaneously broken, from a larger group $G$ into a 
subgroup $H$, anomalies may neither be created nor destroyed, and propagate 
from $G$ to $H$. 
However, the type of the anomaly may change: a local anomaly in $G$ can become 
a global anomaly of $H$. 
For instance, the $SU(2)$ global anomaly in $D=4$ discussed above corresponds 
to a local $SU(3)$ anomaly \cite{Elitzur:1984kr,Okuboetal}.

\section{Constraints from anomalies}
\label{s:constraints}

In this section, we use chiral anomalies to restrict the fermion content of 
(nonsupersymmetric) theories based on various gauge groups in different 
spacetime dimensions. 
More precisely, we shall limit ourself to the study of the possible anomalies 
in every group of  the familiar symmetry breaking sequence\footnote{In 
particular, we do not consider the more string-inspired gauge groups $SO(32)$ 
or $E_8 \otimes E_8$ in $D=10$ or their compactifications to lower 
dimensions \cite{Green:1984sg,Green:1985bx}.}
\begin{mathletters}
\label{E6breaking}
\begin{equation}
E_6 \to SO(10) \to SU(5) \to \SM. \label{E6breakinga}
\end{equation}
For each group, we shall focus on the lowest dimensional representations which 
might be relevant for the Standard Model content, namely the 27 of $E_6$, 
the 16 of $SO(10)$, the $\bar 5$ and 10 of $SU(5)$, and the usual doublets 
and triplets of $\SM$. 
These are related by
\begin{eqnarray}
 & 27 \to 1\oplus 10 \oplus 16 \to 1\oplus (5\oplus\bar 5) \oplus 
(1\oplus\bar 5 \oplus 10),  & \cr
 & \bar 5 \oplus 10 \to ({\cal D}, {\cal L}) \oplus 
({\cal Q}, {\cal U}, {\cal E}), & 
\label{E6breakingb}
\end{eqnarray}
where the second line are respectively one generation of $SU(5)$ and one of 
the Standard Model.
The quantum numbers of the $\SM$ fermions (${\cal Q}$, ${\cal U}$, ${\cal D}$, 
${\cal L}$, ${\cal E}$) are
\begin{equation}
\left(3,2,\frac{1}{3}\right), \ \left(\bar 3,1,\frac{-4}{3}\right), \ 
\left(\bar 3,1,\frac{2}{3}\right), \ (1,2,-1), \ (1,1,2). \label{gen_SM}
\end{equation}
\end{mathletters}
In $D=4k$ dimensions, one may equivalently replace $(\bar 3,1,\frac{-4}{3})$ 
and $(\bar 3,1,\frac{2}{3})$ by $(3,1,\frac{4}{3})$ and $(3,1,\frac{-2}{3})$, 
i.e., replace the left-handed fields ${\cal U}_{L_D}$ and ${\cal D}_{L_D}$ 
with their right-handed charge conjugates $({\cal U}^c)_{R_D}$, 
$({\cal D}^c)_{R_D}$. 
However, since the $(4k+2)$-dimensional charge conjugation does not flip 
chirality, the choices are no longer equivalent in $D=6$ or 10. 

Similar studies have been carried out before, under more restrictive assumptions, 
either without the benefit of the Green-Schwarz mechanism to cancel 
reducible anomalies \cite{Frampton:1984kj}, or using only one representation 
per group to cancel the reducible anomalies \cite{Tosa:1987yf,Tosa:1988bs}.

\subsection{Constraints in $D=6$ dimensions}
\label{s:6D}

As is well known, all groups we mentionned above admit anomaly-free fermion 
content in four dimensions. 
In every case, the anomalies cancel within a generation, and thus they do not 
restrict the number of generations. 
We shall show that the situation is rather different in $D=6$, where anomalies 
yield stronger constraints than in $D=4$. 
We do not impose any {\em a priori} condition on the (six-dimensional) 
chiralities of the representations, which are not constrained by experimental 
results.

\subsubsection{$\SM$ anomalies}
\label{s:SM_6D}

In the case of the group $\SM$, the anomalies which may arise in six 
dimensions are 
\begin{itemize}
\item local gauge anomalies, the only irreducible one being $[SU(3)]^3 U(1)$, 
  and possibly $[U(1)]^4$ if ${\rm Tr}\,Y^2$ vanishes, where $Y$ is the 
  generator of $U(1)$. 
  If ${\rm Tr}\,Y^2 \neq 0$, $[U(1)]^4$ is reducible, as are   $[SU(3)]^4$, 
  $[SU(2)]^4$, $[SU(3)]^2 [SU(2)]^2$, and $[SU(2)]^2 [U(1)]^2$, 
  and they all can be canceled by at most three Green-Schwarz tensors; 

\item mixed anomalies, represented in  Fig.\ \ref{Fig:6Danomalies} (b), 
  where the gauge bosons belong to $SU(3)$, $SU(2)$ or $U(1)$, are reducible, 
  and canceled by the same tensors which are used for the pure gauge 
  anomalies. 
  
\item local gravitational anomalies;

\item global gauge anomalies, since $\Pi_6(SU(3))=Z_6$ and 
  $\Pi_6(SU(2))=Z_{12}$. 
\end{itemize}
All in all, there are four conditions which must be fulfilled: the sums of the 
hypercharges over the $SU(3)$ triplets and antitriplets must vanish. 
Then, there must be as many left- as right-handed fields. 
Finally, using Eq.\ (\ref{global}) with $c_6=2$ for the $SU(2)$ doublets and 
$c_6=1$ for the $SU(3)$ triplets \cite{Bershadsky:1997sb}, we find that the 
numbers of doublets and triplets must satisfy 
\begin{mathletters}
\label{SM_global6D}
\begin{eqnarray}
N(2_{L_6})-N(2_{R_6}) = 0 \mbox{ mod } 6, \label{SM_global6Da}\\
N(3_{L_6})-N(3_{R_6}) = 0 \mbox{ mod } 6.\label{SM_global6Db}
\end{eqnarray}
\end{mathletters}

As mentionned above, the extension of the Standard Model in $D=6$ dimensions 
gives several inequivalent models with different assignments of the quantum 
numbers. 
The consistency of the theory, with a given assignment, has been studied 
previously, under the assumption that local anomalies cancel within a single 
generation \cite{Dobrescu:2001ae}. 
This leads to specific chirality choices for the six-dimensional $\SM$ 
fermions, and to the introduction of an additional singlet in each generation. 
An important result is that it is necessary to have more than one generation, 
in order to cancel the global anomalies \cite{Dobrescu:2001ae}. 
Furthermore, if the generations are identical, i.e., have the same 
chiralities, their number $n_g$ is a multiple of 3, up to an arbitrary number 
of vector-like pairs of generations. 
However, relaxing the requirement that local anomalies cancel in each 
generation, there arise other anomaly-free solutions with three generations.

It is also possible to keep our original quantum number assignment, 
Eq.\ (\ref{gen_SM}). 
With that choice, there is no anomaly-free solution with only one or two 
generations: $n_g\geq 3$. 
There are several solutions with the ``minimal'' three-generation content. 
For instance, one might take three copies of the locally anomaly-free 
generation composed of ${\cal Q}_{L_6}$, ${\cal U}_{R_6}$, ${\cal D}_{R_6}$, 
${\cal L}_{L_6}$, ${\cal E}_{R_6}$, and a right-handed singlet (for the local 
gravitational anomaly). 
A drawback of this solution is that it cannot be embedded in a larger gauge 
group since ${\cal D}$ and ${\cal L}$, for instance, have opposite chiralities.

Another possible solution, with three nonidentical generations, is 
\begin{eqnarray}
\label{SU5_gen}
\left( {\cal Q}_{L_6}, {\cal U}_{L_6}, {\cal D}_{L_6}, {\cal L}_{L_6}, 
{\cal E}_{L_6} \right), \cr
\left( {\cal Q}_{L_6}, {\cal U}_{L_6}, {\cal D}_{L_6}, {\cal L}_{L_6}, 
{\cal E}_{L_6} \right), \\
\left( {\cal Q}_{R_6}, {\cal U}_{R_6}, {\cal D}_{L_6}, {\cal L}_{L_6}, 
{\cal E}_{R_6} \right), \nonumber 
\end{eqnarray}
plus $\SM$ singlets.
In that case, local anomalies do not vanish within a single generation, but 
rather between one generation and two copies of ${\cal D}_{L_6}$ and 
${\cal L}_{L_6}$. 
To obtain three full generations, we used the freedom to add vector-like, 
anomaly-free representations, i.e., ${\cal Q}$, ${\cal U}$, and ${\cal E}$.  
An obvious problem is that the latter could be given Dirac mass terms and be 
decoupled from the low-energy theory. 
A simple, however admittedly inelegant, way to prevent this is to assign some 
discrete symmetry to these extra fields, such as 
${\cal Q}_{R_6} \to -{\cal Q}_{R_6}$  while 
${\cal Q}_{L_6} \to {\cal Q}_{L_6}$; 
it is anyway necessary to implement such a symmetry in order to recover a 
$4D$ chiral theory by compactification on an orbifold. 
 
Finally, note that there are still other 3-generation solutions, as well as 
solutions with for instance $n_g=5$, which are not replications of solutions 
with $n_g=3$.
Nevertheless, we wish to emphasize that there is a feature which does not depend 
on the quantum number assignment. 
If one requires identical generations, then the only anomaly-free solutions 
consist of $n_g = 0$ mod 3 generations, each of which must have no local 
anomaly: each generation only brings 2 or 4 (left- minus right-handed) 
doublets or triplets, while multiples of 6 are necessary, see 
Eqs.\ (\ref{SM_global6D}). 

When this additional condition is not imposed, the number of generations is 
not strictly fixed by the anomaly cancellation requirement. 
This compels us to examine whether even more stringent conditions might be 
derived from larger groups in the sequence Eq.\ (\ref{E6breakinga}).

\subsubsection{$SU(5)$ anomalies in six dimensions}
\label{s:SU5_6D}

It was soon recognized that it might be possible to explain some of the 
arbitrary features of the Standard Model by embedding $\SM$ in a larger 
gauge group \cite{Pati:1973uk}. 
The minimal solution relying on a simple Lie group is the Georgi-Glashow 
$SU(5)$ model \cite{Georgi:1974sy}, in which the Standard Model fermions 
are represented by $(\bar 5\oplus 10)$ generations. 

Let us review the various possible anomalies of a $6D$ $SU(5)$ theory, and, 
first of all, the local gauge anomaly. 
Since $\Pi_7(SU(5))=Z$, a single representation has an irreducible anomaly. 
Taking $D=6$ in Eq.\ (\ref{gaugeDD}), the cancellation condition reads
\begin{equation}
\label{gauge6D}
\sum_{L_6} {\rm STr} (T^a T^b T^c T^d) - 
\sum_{R_6} {\rm STr} (T^a T^b T^c T^d) = 0.
\end{equation}
These traces must be calculated for both representations $\bar 5$ and $10$, 
relating them to the traces over the generators $t^a$ of the fundamental 
$SU(5)$ representation (we follow the notations of \cite{Frampton:1983nr}):
\begin{eqnarray}
\label{Dev_SU5}
\lefteqn{{\rm STr} (T^a_{(k)} T^b_{(k)} T^c_{(k)} T^d_{(k)})  =} & & \cr
& & A_4(5,k)\ {\rm STr}\,(t^a t^b t^c t^d) \cr 
& + &
A_4^{22}(5,k)\ {\rm S}\left({\rm Tr}\,(t^a t^b)\,{\rm Tr}\,(t^c t^d)\right),
\end{eqnarray}
where $k$ labels the representation under study: $k=1$ for the fundamental 
representation 5, $k=2$ for the 10, $k=3$ for the $\overline{10}$, and $k=4$ for 
the $\bar 5$. 
The coefficients are given in Table \ref{tab:A_5k6D}.
Note that the coefficients in the case $k=1$ are trivial. 
\begin{table}[htbp]
\begin{center}
\begin{tabular}{cccc}
& $k$ & $A_4(5,k)$ & $A_4^{22}(5,k)$ \\
\hline
5 & 1 & 1 & 0 \\
10 & 2 & $-3$ & 3 \\
$\overline{10}$ & 3 & $-3$ & 3 \\
$\bar 5$  & 4 & 1 & 0 \\
\end{tabular}
\end{center}
\caption{Coefficients in the symmetrized trace factorization 
Eq.\ (\ref{Dev_SU5}) for the lowest dimension $SU(5)$ representations.}
\label{tab:A_5k6D}
\end{table}

Both traces in the right-hand side of Eq.\ (\ref{Dev_SU5}) are nonvanishing, 
but only the trace over four matrices is irreducible. 
To cancel this first part, one must choose the fermion content of the theory 
appropriately. 
The simplest solution (apart from vector-like generations) made of 
$\bar 5\oplus 10$ representations consists in taking three $(\bar 5)_{L_6}$, 
two $10_{L_6}$, and a $10_{R_6}$: the resulting anomaly is proportional to 
\begin{equation}
3\,A_4(5,4) + 2\,A_4(5,2) - A_4(5,2) = 3 + (-6) - (-3)=0,
\end{equation}
where the minus sign is due to the opposite chirality of the $10_{R_6}$. 
If we consider that a $\bar 5$ and a $10$ form a generation,\footnote{We 
shall comment on this issue in Sec.\ \ref{s:conclusions}.} this means that we 
need three {\em nonidentical} generations to cancel the irreducible part of 
the gauge anomaly. 
Of course, one may add other identical copies of this set of three families: the 
number of generations is $n_g=0$ mod 3.

As stated in Sec.\ \ref{s:local}, the reducible part of the anomaly can be 
canceled through the Green-Schwarz mechanism, by introducing a self-dual, 
antisymmetric tensor \cite{Green:1984sg,Frampton:1985wc}.
The same tensor allows us to also cancel the mixed anomaly.
This latter is proportional to ${\rm Tr}\, (T^a T^b)$, see diagram (b) of 
Fig.\ \ref{Fig:6Danomalies}, and does not vanish with our fermion choice. 

Let us now consider the local gravitational anomaly. 
As recalled above, it vanishes provided the numbers of left- and right-handed 
fields are the same. 
The fermion content imposed by the cancellation of the gauge anomaly consists 
of 5+5+5+10+10 left- and 10 right-handed Weyl fermions. 
In addition, the self-dual antisymmetric tensor contributes for 28 Weyl 
right-handed fermions \cite{Alvarez-Gaume:1984ig}. 
All in all, three additional left-handed fermions, necessarily singlets under 
$SU(5)$, are required to cancel the anomaly.

While there can be local anomalies of every type --- gauge, mixed, and 
gravitational --- unless the fermion content of the theory is carefully chosen, 
the theory cannot be spoiled by the global gauge anomaly, because the 
sixth homotopy group of $SU(5)$ is trivial. 

In conclusion, imposing the absence of anomaly for a chiral $SU(5)$ theory 
in 6 dimensions fixes its gauged fermion content:
\begin{equation}
\label{SU5-6Dmat}
\left( 
  \begin{array}[]{c}
    (\bar 5)_{L_6} \\ 10_{L_6}
  \end{array}
\right) \quad
\left( 
  \begin{array}[]{c}
    (\bar 5)_{L_6} \\ 10_{L_6}
  \end{array}
\right) \quad
\left( 
  \begin{array}[]{c}
    (\bar 5)_{L_6} \\ 10_{R_6}
  \end{array}
\right),
\end{equation}
and requires the introduction of a self-dual antisymmetric tensor and 
three left-handed singlets. 
This solution is not the {\em minimal} anomaly-free solution with $\bar 5$s 
and 10s: we have added a vector-like 10 to obtain three full generations, and 
the remarks following Eq.\ (\ref{SU5_gen}) also apply here. 

What does this solution we propose become when $SU(5)$ is broken into $\SM$?
Given the chirality assignments of Eq.\ (\ref{SU5-6Dmat}), we actually recover 
the three chiral $\SM$ generations of Eq.\ (\ref{SU5_gen}). 
We shall explicitly check that this is indeed an anomaly-free set of three 
generations, although we already know it must be the case since no anomaly can 
have been created when $SU(5)$ was broken. 

As we have seen above, the only irreducible gauge anomaly is $[SU(3)]^3 U(1)$, 
since one easily checks that ${\rm Tr}\,Y^2 \neq 0$.  
This anomaly is proportional to 
\begin{eqnarray}
\lefteqn{2 \times 
\left [ \frac{1}{3}+\frac{1}{3}-\left( \frac{-4}{3}+\frac{2}{3}\right)\right] } & & \cr 
& + & \ \ \left [ -\left(\frac{1}{3}+\frac{1}{3}\right)+
(-1)^2 \left(\frac{-4}{3}\right) - \frac{2}{3} \right] = 0,
\end{eqnarray}
where the factor $(-1)^2$ reflects both right-handed chirality and 
$A_3(3,2)=-1$ for the $\bar 3$. 
The other, reducible anomalies are killed through the Green-Schwarz mechanism: 
the single tensor which was used to cancel the $SU(5)$ anomalies somewhat 
splits into different parts, which in turn cancel the $\SM$ anomalies after 
breaking. 

The $SU(5)$ singlet fermions, which were introduced to cancel the 
gravitational anomaly, are now $\SM$ singlets. 
Since gravity is insensitive to the breaking of other interactions, 
the gravitational anomaly cancellation remains valid.

Global gauge anomalies, on the other hand, depend on the gauge group: while 
there is none in $SU(5)$, both $SU(2)$ and $SU(3)$ can possibly have such 
anomalies.
As seen above, their cancellation requires 0 mod 6 doublets (left- minus 
right-handed) and 0 mod 6 triplets [see Eq.\ (\ref{global})].
Our $SU(5)$-inspired solution satisfies both conditions: 
$N(2_{L_6})-N(2_{R_6})=9-3=6$ and $N(3_{L_6})-N(3_{R_6})=9-3=6$. 
From the point of view of $\SM$, this is the condition which suggests some 
restriction on the number of generations. 
In $SU(5)$, the condition comes from the irreducible part of the gauge 
anomaly. 
This is a striking example of the change of nature of an anomaly when a group 
is broken.

Therefore, the anomaly-free three-generation $SU(5)$ theory becomes an 
anomaly-free $\SM$ theory when the symmetry is broken, as expected. 
On the other hand, all other anomaly-free, six-dimensional $\SM$ theories do 
not originate from a $SU(5)$ theory. 
This is for instance the case of the solution with 3 identical generations we 
mentionned in the discussion on $\SM$ anomalies, since the ${\cal D}$ and 
${\cal L}$ have opposite chiralities, and cannot come from a single $\bar 5$. 

We have summarized in table \ref{tab:SU5} the different anomalies which can 
affect $SU(5)$ and $\SM$ theories in six dimensions. 
Note that while there is no global anomaly anomaly in $SU(5)$, there is one in 
the Standard Model group, which automatically vanishes for a $\SM$ matter 
content which comes from $SU(5)$. 

\begin{table}[htbp]
\begin{center}
\begin{tabular}{lcc}
$D=6$ & $SU(5)$ & $\SM$ \\
\hline
pure gauge & yes & yes \\
mixed & yes & yes \\
gravitational & yes & yes \\
global & no & yes \\
\hline
\end{tabular}
\end{center}
\caption{Possible $SU(5)$ and $\SM$ anomalies in 6 dimensions.}
\label{tab:SU5}
\end{table}

\subsubsection{Six-dimensional $SO(10)$ and $E_6$ anomalies}
\label{s:S010,E6_6D}

As we have just shown, a six-dimensional $SU(5)$ theory is anomaly-free only 
if the number $n_g$ of chiral generations is a multiple of 3, with specific 
chirality assignments for the various $\bar 5$ and $10$ which yield the matter 
content of the Standard Model, plus additional $SU(5)$ singlets.
For the economic solution $n_g=3$, there are three such singlets. 

A similar result was obtained in \cite{Dobrescu:2001ae}, where each chiral 
generation must be added a $\SM$ singlet to cancel the gravitational anomaly.
In both cases, it is necessary to introduce as many gauge singlets as there are 
chiral generations, although it should be noted that the underlying reasons 
are different. 
In \cite{Dobrescu:2001ae}, the number of extra fermions is necessarily equal 
to the number of chiral families, since the local anomalies are required to 
vanish within each generation. 
On the other hand, in the present paper, we do not impose this condition, and 
we still have to add 3 $SU(5)$ singlets to our 3 chiral generations.

This coincidence naturally leads to the idea that both models could be 
embedded in more fundamental theories based on a larger symmetry group. 
The first obvious candidate, which unifies the 15 Weyl fermions of a given 
generation of the Standard Model with a $SU(5)$ or $\SM$ singlet in a single 
representation, is the orthogonal group $SO(10)$, with its spinorial 16
representation \cite{refSO10}. 
Next, we shall consider the case of the exceptional group $E_6$. 

The case of $SO(10)$ is rather different from the previous groups we 
considered. 
Since $\Pi_7(SO(10))=Z$, there must be an irreducible anomaly. 
Indeed, ${\rm Tr}\,(T^a)^4=16$ for any generator of the 16 spinor, and the 
local pure gauge anomaly, Eq.\ (\ref{gauge6D}), has a nonvanishing irreducible 
part \cite{Tosa:1987yf,Hebecker:2001jb}. 
Therefore, a chiral six-dimensional $SO(10)$ theory cannot be anomaly-free 
if it only contains copies of the 16 representation.

It was quite obvious from the beginning that the 3-generation, anomaly-free 
$SU(5)$ solution Eq.\ (\ref{SU5_gen}) cannot be trivially promoted to an 
$SO(10)$ model. 
As a 16 of $SO(10)$ transform as $\bar 5\oplus 10 \oplus 1$ under $SU(5)$, the 
$(\bar 5)_{L_6}$ and $10_{R_6}$ of the third generation cannot originate from 
a single 16, which is either left- or right-handed. 
But since we have shown that a $16$ of $SO(10)$ is always anomalous in six 
dimensions, it cannot either yield a consistent $\SM$ theory. 

To cancel this anomaly, one should either add another 16 with opposite 
chirality --- but this amounts to losing the chirality of the theory ---, or 
add some other, new matter field, which spoils the simplicity researched when 
embedding $\SM$ or $SU(5)$ in $SO(10)$. 
For example, one might consider two left-handed and a right-handed 16, plus 
a left-handed 10. 
The $16_{R_6}$ and a $16_{L_6}$ form a vector-like pair, and are therefore 
anomaly-free.
The irreducible parts of the pure gauge anomalies of the remaining $16_{L_6}$ 
and the $10_{L_6}$ cancel \cite{Tosa:1987yf,Hebecker:2001jb}, as will be 
obvious when we discuss $E_6$. 
The other local anomalies are in a sense harmless, since they can be canceled 
by the Green-Schwarz mechanism (reducible anomalies) or by adding gauge 
singlets (gravitational anomaly). 
Furthermore, there is no global gauge anomaly, because $\Pi_6(SO(10))=0$. 

When $SO(10)$ is broken into $SU(5)$, the 16 of this anomaly-free solution 
transform as in Eq.\ (\ref{E6breakingb}), while the $10_{L_6}$ yields 
$(5 \oplus \bar 5)_{L_6}$. 
All in all, we recover the fermion content of Eq.\ (\ref{SU5_gen}), with in 
addition a pair $5_{L_6} \oplus (\bar 5)_{R_6}$, in which the irreducible 
anomalies cancel, see Table \ref{tab:A_5k6D}. 
This pair can be decoupled from the low energy spectrum by a Dirac mass term. 

Let us now consider $E_6$. 
In that group, the basic, nonfactorizable traces are the traces over products 
of 2, 5, 6, 8, 9, and 12 generators \cite{Tosa:1987yf}. 
Therefore, the local gauge anomaly is factorizable in $D=6$. 
There only remains a reducible part, which can be canceled through the 
Green-Schwarz mechanism, which also protects the theory against mixed 
anomalies.
Similarly, the gravitational anomalies of a single 27 can be canceled by 
adding singlet fermions with the appropriate chirality. 
Since the sixth homotopy group of $E_6$ is trivial, $E_6$ has no global gauge 
anomaly in six dimensions.
Thus, a six-dimensional $E_6$ theory, with a Green-Schwarz tensor and 
singlet fermions, is anomaly-free, whatever the number of 27s of the theory.

When the symmetry is broken, the absence of anomaly for a single 27 of $E_6$ 
is transmitted to the $SO(10)$ representation $16\oplus 10 \oplus 1$. 
The singlet obviously has no gauge anomaly, and we recover the fact that the 
irreducible anomalies of a $16$ and a $10$ with identical chiralities cancel. 

Going further in the sequence of Eq.\ (\ref{E6breakinga}), the same idea 
suggests that the irreducible parts of the pure gauge anomalies of a $5$, 
two $\bar 5$, and a $10$ of $SU(5)$ with identical chiralities should cancel. 
They do indeed, as show the values of $A_4(5,1)$, $A_4(5,4)$, and $A_4(5,2)$ 
in Table \ref{tab:A_5k6D}. 
This solution with three $5$ or $\bar 5$ and a $10$ is the minimal $SU(5)$ 
fermion content free of irreducible anomalies in $D=6$. 
However, it does not contain all observed fermions of the Standard Model, 
which require at least 3 copies of this basic, anomaly-free building block. 
This then yields too many fermions, while there exists a more economical 
solution, Eq.\ (\ref{SU5_gen}).
As mentionned before, the latter consists in fact of a left-handed copy of the 
minimal $(\bar 5 \oplus \bar 5 \oplus \bar 5 \oplus 10)$ content, plus a 
vector-like, anomaly-free pair $10_{L_6}\oplus 10_{R_6}$.

\subsection{Constraints from anomalies in $D=8$}
\label{s:8D}

An important difference which characterizes the study of anomalies in eight 
dimensions is the absence of local gravitational anomalies, since the 
spacetime is $4k$-dimensional. 
This removes a condition which automatically led to the introduction of 
singlet fermions, which we shall now be able to discard, unless we wish to 
introduce some to cancel the global gravitational anomaly.

\subsubsection{$\SM$ anomalies in $D=8$}
\label{s:SM_8D}

When the spacetime dimension $D \geq 8$, theories based on the Standard Model 
group are less constrained than in lower dimensions. 
This is because almost all local gauge anomalies are now reducible: for 
$SU(3)$ and $SU(2)$, both traces
\begin{equation}
{\rm STr} \left( T^a T^b T^c T^d T^e\right) \label{gauge8D}
\end{equation}
and
\begin{equation}
{\rm STr} \left( T^a T^b T^c T^d T^e T^f\right)\label{gauge10D}
\end{equation}
are factorizable, and therefore the corresponding anomalies can be canceled by 
Green-Schwarz tensors. 
So can be the $[U(1)]^5$ and $[U(1)]^6$ anomalies in most cases. 
Finally, gauge anomalies with bosons from different groups are always 
reducible. 

Most mixed gauge-gravity anomalous diagrams are also reducible, with the 
exception of the $D=8$ pentagonal diagram with four gravitons and a $U(1)$ 
boson. 
This diagram yields a contribution proportional to ${\rm Tr}\,Y$. 
Thus, in $D=8$, the gauged fermions of a $\SM$ theory must satisfy the 
condition ${\rm Tr}\,Y=0$.

In eight dimensions, a simple solution consists in building a generation with 
left-handed versions of the ${\cal Q}$, ${\cal U}$, and ${\cal D}$ of 
Eq.\ (\ref{gen_SM}), and right-handed ${\cal L}$ and ${\cal E}$.
A single generation is then free of the only irreducible local anomaly, since 
${\rm Tr}\,Y=0$. 
In that particular case, ${\rm Tr}\,Y^3\neq 0$, so that the $[U(1)]^5$ anomaly 
is reducible \cite{Tosa:1988bs}.\footnote{On the other hand, in the case of a 
generation where all the fields of Eq.\ (\ref{gen_SM}) are left-handed, which 
satisfies ${\rm Tr}\,Y=0$ as well, the chiralities are such that 
${\rm Tr}\,Y^3$ also vanishes, and the pure gauge anomaly $[U(1)]^5$ is not 
reducible.}
Besides, such a generation contains $N(3_{L_8})-N(3_{R_8})=0$ triplets and 
two left-handed doublets. 
It therefore automatically satisfies the conditions, Eq.\ (\ref{global}), 
necessary to cancel the $SU(3)$ and $SU(2)$ global anomalies, due to 
$\Pi_8(SU(3))=Z_{12}$ and $\Pi_8(SU(2))=Z_2$. 
The only remaining constraint comes from the global gravitational anomaly, 
which requires either an even number of generations, or the introduction of an 
additional spin $\frac{1}{2}$ singlet fermion. 

Therefore, a single $\SM$ generation, Eq.\ (\ref{gen_SM}), with appropriate 
chirality choices, is anomaly-free, with the help of Green-Schwarz tensors to 
cancel the reducible anomalies and a Weyl fermion to protect the theory 
against the global gravitational anomaly. 
Anomalies give no restriction on the number of such generations.

\subsubsection{$SU(5)$ in eight dimensions}
\label{s:SU5_8D}

The anomalies of $SU(5)$ in $D=8$ are rather similar to the case $D=6$, since 
$\Pi_9(SU(5))=Z$ means that there is an irreducible local gauge anomaly, while 
$\Pi_8(SU(5))=0$ guarantees the absence of global gauge anomaly. 

The pure gauge anomaly Eq.\ (\ref{gauge8D}) is non-factorizable, so that the 
irreducible part must be canceled by the fermion content. 
The expansion in terms of symmetrized traces over the generator of the basic 
representation involves the coefficients $A_5$ given in Table 
\ref{tab:A_5k8D}. 
Given the values of $A_5(5,2)$ and $A_5(5,4)$, the most economic choice 
consists in taking $n_g=11$ generations ($\bar 5 \oplus 10$), with appropriate 
chiralities: 
five ($(\bar 5)_{L_8}\oplus 10 _{L_8}$) and six 
($(\bar 5)_{L_8}\oplus 10 _{R_8}$).

\begin{table}[htbp]
\begin{center}
\begin{tabular}{cccc}
$D=8$ & $k$ & $A_5(5,k)$ & $A_5^{32}(5,k)$ \\
\hline
5 & 1 &1 & 0 \\
10 & 2 & $-11$ & 10 \\
$\overline{10}$ & 3 & $11$ & $-10$ \\
$\bar 5$  & 4 & $-1$ & 0 \\
\end{tabular}
\end{center}
\caption{Coefficients in the symmetrized trace factorization for the lowest 
dimensions $SU(5)$ representations in 8 dimensions.}
\label{tab:A_5k8D}
\end{table}

The remaining, reducible part of the pure gauge anomaly, as well as the mixed 
anomaly, can be canceled through the Green-Schwarz mechanism. 
As announced above, there is no global gauge anomaly. 

Finally, the cancellation of the global gravitational anomaly necessitates an 
even number of spin $\frac{1}{2}$ Weyl fermions. 
This requires the introduction of an odd number of singlet fermions. 
If one prefers not to add sterile matter, the global gravitational anomaly 
rules out the $n_g=11$ solution which cancels the local gauge anomaly. 
The ``minimal'' anomaly-free solution then consists of twice the 11-generation 
solution, i.e., requires $n_g=22$ reducible ($\bar 5 \oplus 10$) generations. 

When $SU(5)$ breaks into $\SM$, the 11-generation solution remains 
anomaly-free: 
the condition ${\rm Tr}\,Y$ is satisfied, since $Y$ is a generator of $SU(5)$; 
the eleven ($\bar 5 \oplus 10$) yield $N(3_{L_8})-N(3_{R_8})=-12$ triplets and 
$N(2_{L_8})-N(2_{R_8})=8$ doublets, so that there is no global gauge anomaly. 

On the other hand, the $D=8$ anomaly-free $\SM$ generation we found previously 
does not originate from $SU(5)$ because of, for example, the opposite 
chiralities of the ${\cal D}_{L_8}$ and ${\cal L}_{R_8}$. 
As in the case of six dimensions, an eight-dimensional $SU(5)$ theory is more 
constrained than a theory based on $\SM$.

\subsubsection{$SO(10)$ and $E_6$ in eight dimensions}
\label{s:SO10,E6_8D}

An $SO(10)$ model suffers from the same problems in $D=8$ as in six 
dimensions. 
Consider, once again, a single 16. 
The mixed gauge-gravitational anomaly involves traces over either one or three 
antisymmetric generators, and therefore vanishes.
The local gauge anomaly Eq.\ (\ref{gauge8D}), on the other hand, is 
nonvanishing for a 16: the trace over five $SO(10)$ generators is proportional 
to the totally antisymmetric Levi-Civita tensor with 10 indices, see Appendix 
\ref{s:Tr(SO10)^5}.
The possible remedies to cure this anomaly are the same as in six dimensions: 
either make the theory vector-like, or add matter fields in different $SO(10)$
representations. 
In the latter case however, it becomes necessary to take care of the global 
gauge anomaly, arising from the homotopy group $\Pi_8(SO(10))=Z_2$, which will 
constrain the number of generations as in the case of $\SM$ in 6 dimensions.
In addition, there must be an even number of spin $\frac{1}{2}$ fermions, to 
cancel the global gravitational anomaly. 

Using the same reasoning as in $D=6$ dimensions, the eight-dimensional 
anomaly-free $SU(5)$ theory with $n_g=11$ generations we have encountered 
above does not come from an $SO(10)$ theory, because of the 
$(\bar 5)_{L_8}\oplus 10_{R_8}$ generations. 
There is therefore no contradiction with the impossibility of building 
anomaly-free $SO(10)$ theories with the 16 representation. 

To study the possible anomalies of a 27 in $D=8$, we shall not follow the same 
procedure as in the six-dimensional case. 
When $E_6$ is broken into $SO(10)$, the 27 transforms following 
Eq.\ (\ref{E6breakingb}). 
As always,  no anomalies are created in this symmetry breaking. 
We have seen above that the 16 of $SO(10)$ is anomalous. 
However, a 10 is anomaly-free, as can be checked by calculating the trace over 
the product of five real generators, Eq.\ (\ref{gauge8D}). 
Therefore, a single 27 of $E_6$ has an irreducible gauge anomaly in eight 
dimensions. 
Another, less pedagogical way of seeing this anomaly consists in noticing that 
$\Pi_9(E_6)=Z$. 

The same reasoning allows us to check without calculation that a single 16 of 
$SO(10)$ is anomalous in 6 or 8 dimensions. 
Since the $\bar 5$ and $10$ have local anomalies which do not cancel (this is 
precisely why three or eleven generations are required), a parent 16 cannot be 
anomaly-free. 

On the other hand, $E_6$ has no mixed anomaly, since the 27 has no pure gauge 
anomaly in $D=4$ dimensions. 
There is no global gauge anomaly either: $\Pi_8(E_6)=0$. 
Finally, the global gravitational anomaly can as always be canceled through 
the introduction of an extra singlet. 

Nonetheless, a chiral theory $E_6$ with fermions only in the 27 representation 
is anomalous in $D=8$ dimensions, due to the irreducible local gauge anomaly, 
which also spoils chiral eight-dimensional $SO(10)$ theories.

\subsection{Anomalies in $D=10$}
\label{s:10D}

\subsubsection{$\SM$ in ten dimensions}
\label{s:SM_10D}

As we have announced in Sec.\ \ref{s:8D}, in $D=10$, there is in most cases 
no constraint on the possible $\SM$ gauged fermions arising from local 
anomalies: only the $[U(1)]^6$ anomaly might be irreducible, under specific 
conditions. 
One may for example build a generation with left-handed versions of all 
fields. 
In that case, ${\rm Tr}\,Y^2\neq 0$, so that the $[U(1)]^6$ anomaly is 
reducible \cite{Tosa:1988bs}. 
Both $SU(3)$ and $SU(2)$ homotopy groups are non-trivial: 
$\Pi_{10}(SU(3))=Z_{30}$ and $\Pi_{10}(SU(2))=Z_{15}$, and there might be 
global gauge anomalies. 
Since the number of triplets in a generation is even, a solution with $n_g=15$ 
generations is obviously free of these anomalies. 
However, we shall see that information derived from $SO(10)$ will allow us to 
improve this result: even a single generation will be found to have no global 
anomalies.

\subsubsection{$SU(5)$ anomalies in $D=10$}
\label{s:SU5_10D}

Since $SU(5)$ is a rank 4 group, there are four Casimir operators and as many 
basic traces, over products of 2, 3, 4, and 5 generators. 
Indeed, the trace of the product of six generators can be factorized in terms 
of these basic traces: in the case of the fundamental representation 
\cite{Frampton:1983nr},
\begin{eqnarray}
\lefteqn{{\rm STr}\left(t^a t^b t^c t^d t^e t^f \right)=} & & \cr
& & \frac{3}{4} {\rm S}\left[ {\rm Tr}\left(  t^a t^b t^c t^d \right) 
{\rm Tr} \left(  t^e t^f \right) \right] \cr
& - & \frac{1}{8} {\rm S}\left[ {\rm Tr}\left(  t^a t^b \right) 
{\rm Tr} \left(t^c t^d \right) {\rm Tr} \left(  t^e t^f \right) \right] \cr
& + & \frac{1}{3} {\rm S}\left[ {\rm Tr}\left(  t^a t^b t^c \right) 
{\rm Tr} \left(  t^d t^e t^f \right) \right].
\end{eqnarray}

Therefore, after an expansion in traces over the generators of the fundamental 
representation, the pure gauge anomaly Eq.\ (\ref{gauge10D}) is fully 
reducible, and so is the mixed anomaly. 
Accepting the introduction of the 2-, 4-, and 6-forms of the Green-Schwarz 
mechanism, these anomalies do not constrain the gauged fermion content, while 
it is impossible to cancel all these anomalies with only an appropriate choice 
of matter content \cite{Frampton:1983nr}. 
The gravitational anomaly does not vanish either, but can be canceled easily 
by singlet fermions. 
Therefore, nothing constrains the chiralities of the $\bar 5$ and $10$ which 
constitute a generation, and we are free to choose them both left-handed, in 
order to recover the $\SM$ solution we proposed above. 

The only stringent condition comes from the global gauge anomaly, due to 
$\Pi_{10}(SU(5))=Z_{120}$. 
Knowing the actual condition requires some knowledge of the coefficients 
$c_{10}$ [see Eq.\ (\ref{global})] for the representations of $SU(5)$ in 10 
dimensions. 
In any case, models with 120$k$ left-handed generations are anomaly free, 
although more ``economical'' solutions exist, as we show hereafter.

\subsubsection{$SO(10)$ and $E_6$ in ten dimensions}
\label{s:SO10,E6_10D}

Let us now consider $SO(10)$ in $D=10$ dimensions. 
A single 16 representation suffers from irreducible gauge anomalies, since 
$\Pi_{11}(SO(10))=Z$, so that chiral theories containing only this 
representation are no more consistent than in $D=6$ or $D=8$.
Furthermore, for locally anomaly-free theories with an extended representation 
content, there is also a possible global anomaly due to 
$\Pi_{10}(SO(10))=Z_4$. 

Consider for example a $(16\oplus 10 \oplus 10)$ representation. 
The irreducible part of the local gauge anomaly is zero, thanks to the choice 
of fermion content \cite{Tosa:1987yf}. 
The remaining, reducible part of the anomaly and the mixed anomaly are 
canceled by Green-Schwarz tensors. 
One can easily cancel the gravitational anomaly with sterile fermions. 
Finally, four copies of $(16 \oplus 10 \oplus 10)$ representations with 
identical chiralities will automatically have no global anomaly. 

Let us break $SO(10)$ into $\SM$; the resulting theory is necessarily 
anomaly-free. 
Each $(16 \oplus 10 \oplus 10)_{L_{10}}$ yields eight left-handed triplets or 
antitriplets and as many doublets. 
Since the theory with broken symmetry has no anomaly, it is in particular free 
of global anomalies. 
Therefore, the conditions $N(3_{L_{10}})-N(3_{R_{10}})=32$ and 
$N(2_{L_{10}})-N(2_{R_{10}})=32$ are sufficient to cancel the $SU(3)$ and 
$SU(2)$ global anomalies. 
Inserting these results in Eq.\ (\ref{global}), we now have further 
information on the coefficients $c_{10}$ for the doublets and triplets:
both $c_{10}(2)$ and $c_{10}(3)$ are multiple of 15. 
This allows us to significantly improve our condition for $\SM$ in ten 
dimensions: 
instead of 15 generations, we see that a single one is already anomaly-free. 

Therefore, in $D=10$, $\SM$ theories with any number of such generations are 
anomaly-free, and anomalies do not constrain $n_g$. 

Since an $\SM$ generation of left-handed fermions is anomaly-free, so is an 
$SU(5)$ generation: a single $(\bar 5 \oplus 10)_{L_{10}}$ is already free of 
the potential global anomaly, while we only knew that a sufficient choice was 
$n_g=0$ mod 120. 
This means that the coefficients $c_{10}(\bar 5)$ and $c_{10}(10)$ are both 
multiple of 120. 
As in the case of $SU(2)$ and $SU(3)$, it is not necessary to actually 
calculate the coefficients to obtain interesting information. 

We have just seen that a single $(\bar 5 \oplus 10)_{L_{10}}$ is anomaly-free 
in $SU(5)$ , while we mentionned previously that a 16 of $SO(10)$ is 
anomalous. 
This means that the $SU(5)$ theory cannot come from $SO(10)$, although the 
explanation is more subtle than in $D=6$ or 8. 
The symmetry breaking sequence Eq.\ (\ref{E6breakinga}) could be further 
specified: in fact, $SO(10)$ breaks into $SU(5) \otimes U(1)$, so that the 
$\bar 5$ and 10 of $SU(5)$ also carry some $U(1)$ charge (3 and $-1$ 
respectively \cite{Slansky:1981yr}), which gives rise to a new anomalous 
diagram $[SU(5)]^5 U(1)$. 
The contribution of this diagram, which is absent in a pure $SU(5)$ theory, 
does not vanish, so that the $SU(5) \otimes U(1)$ theory is not anomaly-free, 
whatever the number of generations, as the parent $SO(10)$ theory. 

The ten-dimensional pure gauge anomaly of the 27 of $E_6$ can be checked in 
the same way as in $D=8$. 
Both 10 and 16 representations of $SO(10)$ are anomalous, and the irreducible 
parts of their anomalies do not cancel \cite{Tosa:1987yf}. 
Therefore, the 27 is necessarily anomalous, as also shows the homotopy group 
$\Pi_{11}(E_6)=Z$. 
Although other anomalies either can be canceled (in the case of the mixed and 
gravitational anomalies) or vanish [$\Pi_{10}(E_6)=0$, hence there is no 
global gauge anomaly], there is no way of obtaining an anomaly-free $E_6$ 
theory with only copies of the 27, as in $D=8$. 

In summary, in $D=10$ it is possible to build chiral $\SM$ or $SU(5)$ theories 
with any number of generations, but they do not come from $SO(10)$ or $E_6$ 
since a 16 or a 27 are always anomalous.

\section{Anomaly-free $SU(5)$ model in 6 dimensions}
\label{s:model}

In this section, we investigate some of the properties of a six-dimensional 
model implementing the anomaly-free three-generation $SU(5)$ fermion content 
of Eq.\ (\ref{SU5_gen}). 
A salient feature of these three generations is of course their non-identity. 
We shall first investigate whether this leads to specific characteristics at 
tree level (Sec.\ \ref{s:model1}).
Then, we review in Sec.\ \ref{s:6to4D} some of the possible ways of going from 
$SU(5)$ in $D=6$ to the Standard Model in $D=4$.

\subsection{Basic properties of models with three nonidentical $SU(5)$ 
generations}
\label{s:model1}

As we have seen in Sec.\ \ref{s:constraints}, the most stringent constraints 
imposed by anomalies on the fermion content affect six- and eight-dimensional 
$SU(5)$ theories. 
The most promising case is obviously $D=6$, which hints at the necessary 
existence of three generations, as in the Standard Model. 

A nice feature of the $SU(5)$ fermion content in six dimensions is the 
difference between the third generation ($\bar 5_{L_6} \oplus 10_{R_6}$) and 
the other two. 
This might explain why the third Standard Model generation is so much heavier 
than the lightest two, and we shall assume that the 
$\bar 5_{L_6} \oplus 10_{R_6}$ yields, after breaking, the top, bottom, 
$\tau$, and $\nu_\tau$, although our discussion will not depend on this 
assumption. 

Following this idea, let us examine the possible six-dimensional mass terms 
which could be given to the $SU(5)$ fermions. 
The so-called minimal symmetry breaking scheme for $SU(5)$ in $D=4$ involves 
two Higgs fields in the 5 and 24 representations. 
The former is used to break $SU(5)$ and cannot give rise to a mass term. 
In fact, we shall see in Sec.\ \ref{s:6to4D} that this field is not even 
necessary to break the symmetry, if this is done through the compactification 
on an orbifold. 

In opposition, the 5 Higgs field may yield $SU(5)$ singlet terms, thanks to 
the tensor product decompositions \cite{Slansky:1981yr}
\begin{eqnarray}
\label{SU5massterms}
  5  \otimes  {\bar 5} & = & 1\oplus24 \cr
  \bar 5 \otimes \bar 5 & = & \overline{10} \oplus \overline{15} \cr
  10 \otimes 10 & = & \bar 5 \oplus 45 \oplus 50 \cr
  \bar 5 \otimes 10 & = & 5 \oplus \overline{45}. 
\end{eqnarray}
{\em A priori}, $SU(5)$ invariance allows the construction of mass terms 
with two $10$ or with a $\bar 5$ and a $10$, independently of the spacetime 
dimension or the fermion chiralities. 
However, these two ingredients do influence mass terms, since the latter must 
be ($D$-dimensional) Lorentz invariant. 

The chirality of the $10$ of the third generation is opposite to that of all 
other $\bar 5$ and $10$, so that a term involving this $10_{R_6}$ and any 
other fermion can be Lorentz invariant.  
However, in order to have three light generations in $4D$, such a term should 
be forbidden. [See the discussion following Eq. (\ref{SU5_gen}).]
Furthermore, it is not possible to build a Lorentz invariant mass term with 
only left-handed fermions in $D=6$. 
Hence,  at tree level, most, but not all, of the fermions will remain massless 
after symmetry breaking.
Note that this feature is due to  the different chirality assignment of 
the different generations in $6D$, and not only to the structure of the gauge 
group. 
Of course, the situation will be modified by radiative corrections, which can 
generate mass terms after dimensional reduction to $4D$. 
The rather restricted number of  couplings will also affect the CKM
matrix and one might hope to obtain in a natural way some {\sf CP} violation. 
In addition, compactification scheme leading to different localization 
patterns of the generations may further enrich the discussion. 
It would be interesting to see whether such a scheme, be it grand unified or not, 
 could give rise to non-trivial mass textures.  
This would relate the hierarchy of fermion masses in the Standard Model to 
the cancellation of anomalies in $6D$.

Before we turn to the issue of symmetry breaking and compactification, let 
us mention another prediction of the theory. 
In any Grand Unified Theory, the weak mixing angle can be related to the 
traces over two generators of the group \cite{Langacker:1981js}: 
\begin{equation}
  \sin^2 \Theta_W = \frac{{\rm Tr}\,T_3^2}{{\rm Tr}\,Y^2}. 
\end{equation}
In four dimensions, if all the content of a single family is incorporated in 
one or more irreducible representations of the GUT group, the prediction does 
not depend on the group. 
These GUT representations could contain other states, but the latter have to 
be singlets under $\SM$, and the result is thus the same for $SU(5)$ and 
$SO(10)$, namely $\sin^2 \Theta_W=3/8$. 
Nonetheless, radiative corrections have to be included, which strongly depend 
on the nature of the GUT group. 

In our six-dimensional case, the predicted value for the first two families is 
of course the same as in four dimensions, since the chiralities are identical. 
In $D=4$, one could use charge conjugation to flip some of the chiralities
thereby changing the corresponding signs, and the result would not change 
thanks to the square powers: this shows without calculation that the value is 
identical for the $\bar 5_{L_6} \oplus 10_{R_6}$, despite the opposite 
chirality of the 10. 
Therefore, the value in our six-dimensional $SU(5)$ is also 
$\sin^2 \Theta_W=3/8$.

\subsection{Dimensional reduction and $SU(5)$ breaking}
\label{s:6to4D}

The combination of extra dimensions and GUT models brings together the issues 
related with both aspects. 
Thus, one should investigate the running of the Standard Model coupling 
constants, proton decay \cite{Langacker:1981js,Kobakhidze:2001yk}, the 
presence of monopoles, the hierarchy problem... 
The actual features of the various problems depend heavily on the localization 
and compactification mechanisms. 
Although we shall not go into detail in the present paper, we would like to 
make a few comments and suggestions to show that our approach is not grossly 
ruled out. 

Starting from a $SU(5)$ model in $D=6$ dimensions, there are several paths 
towards $\SM$ in four dimensions. 
Two different steps are required, which may be simultaneous or not: $SU(5)$ 
must be broken, and the six-dimensional theory must be reduced into a 
four-dimensional theory. 
The second operation should also involve some chirality selection, since the 
fermions of the Standard Model have definite, left-handed chiralities. 

In the absence of a reliable mechanism to suppress the proton decay, a 
compactification scheme which allows $SU(5)$ to survive in four dimensions 
should likely be discarded. 

There has been recently a growth of interest in five-dimensional $SU(5)$  
models, either supersymmetric or not, with an orbifold $S^1/Z_2$ or 
$S^1/(Z_2\otimes Z'_2)$ \cite{Hebecker:2001jb,SU5-5D,Kobakhidze:2001yk}, 
as well as in $D=6$ $SU(5)$ or $SO(10)$ models with an orbifold 
$T^2/(Z_2\otimes Z'_2)$ \cite{ref6D,Hall:2001xr}. 
In these studies, Standard Model fermions are totally confined to an orbifold 
fixed point, corresponding to our four-dimensional brane. 
The boundary conditions on the gauge bosons can be fixed appropriately so as 
to break $SU(5)$ to $\SM$: only the zero modes of the Standard Model gauge 
bosons have a non-zero value at the fixed point. 
This is an efficient way to reduce proton decay probability, since there is no 
overlap between the quarks and leptons and the $SU(5)$-specific, 
{\sf B}-violating bosons. 

In our six-dimensional case, if the Standard Model fields are strongly 
localized, for instance with a vortex \cite{Frere:2001dc,Callan:1985sa}, we 
may then benefit from the same effect.
In addition, it has been showed that the localization of chiral fermions can 
enhance, through loop effects, the localization of the zero modes of the gauge 
bosons to which the fermions are coupled \cite{Dvali:2001rx}. 
This further increases the suppression of unwanted $SU(5)$ effects. 
Besides, possible Standard Model contributions to the proton decay 
may be suppressed by a residual spacetime symmetry, relic of the
six-dimensional Lorentz invariance after compactification 
\cite{Appelquist:2001mj}.  
This mechanism is however model dependent.  
Whether a similar effect could suppress {\sf B}-violating processes in the 
$SU(5)$ model discussed here is an open question. 

An issue related with the localization of the Standard Model fields is the 
size of the extra dimensions. 
If the fields are strongly localized, as seems to be necessary to avoid an 
important proton decay, then the extra dimensions might be large, with radii 
of the order of (10 TeV)$^{-1}$. 
This is also the Grand Unification energy scale, where the Standard Model 
couplings unify \cite{Dienes:1999vg}. 
The unusual chirality of our third generation does not modify the actual value 
much. 

To be fair, one must admit that, although the compactification on an orbifold 
has many advantages, since it allows both chirality selection and symmetry 
breaking, it is in no way predictive. 
First, it might lead to an anomalous four-dimensional theory if the $4D$ 
chiralities are not properly selected.  
Then, one could hope that the constraint of the absence of anomaly in four 
dimensions would only leave a single possibility, the Standard Model. 
Unfortunately, this is obviously not the case. 
As is clear from (\ref{SU5_gen}), with an appropriate choice of orbifold, 
we could as well get a single (anomaly-free) generation in $D=4$, plus 
vector-like fermions which could then be decoupled from the low energy 
spectrum. 
Therefore, even though anomalies might give an explanation of the existence 
of three generations, they are not restrictive enough so as to permit only one 
particular fermion content after dimensional reduction. 

Another potentially interesting feature of our three-generation $SU(5)$ 
solution is the necessary existence of three extra singlets which, after 
compactification, will give rise to Kaluza-Klein towers of sterile states 
\cite{Dienes:1999sb}.
In turn, these states can give masses to the light neutrinos. 

Finally, the compactification does not wholly suppress the Green-Schwarz 
tensor which was required to cancel reducible $SU(5)$ and mixed anomalies. 
There remain some of the tensor components, with axion-like couplings 
\cite{Green:1987mn}. 
In an elaborate model, far beyond the scope of the present paper, these latter 
could be used to suppress the unobserved strong {\sf CP} violation 
\cite{Fabbrichesi:2001fx}

\section{Conclusions}
\label{s:conclusions}

As we have shown in this paper, anomaly cancellation restricts not only 
the chiral fermion content, but also the possible Yukawa couplings of Grand 
Unified Theories propagating in extra dimensions. 

\begin{table}[htbp]
\begin{center}
\begin{tabular}{lccc}
& $D=6$ & $D=8$ & $D=10$ \\
\hline
Standard Model & $n_g \geq 2$ & $n_g$ arbitrary & $n_g$ arbitrary \\
$SU(5)$ & $n_g=3k$ & $n_g=11k$ & $n_g$ arbitrary \\
$SO(10)$ & \multicolumn{3}{c}{no solution with only copies of the 16} \\
$E_6$ & $n_g$ arbitrary & \multicolumn{2}{c}{no solution with 27 only}\\
\end{tabular}
\end{center}
\caption{Constraints on the gauged chiral fermion content of various theories 
in different spacetime dimensions.}
\label{tab:summary}
\end{table}

Our results are summarized in Table \ref{tab:summary}. 
For instance, we have shown that $SO(10)$ is not a good Grand Unification 
candidate, be it in $D=6$, 8 or 10, if the Standard Model fermions are 
represented by the 16. 
One needs to invoke other matter fields, beyond the Standard Model, in some 
other representation (see \cite{Hall:2001xr} for a similar discussion in the 
supersymmetric context). 
But this might mean that the relevant group is $E_6$, rather than $SO(10)$. 
On the other hand, six-dimensional $SU(5)$ theories can be anomaly-free, 
provided the matter content is very finely tuned: if a generation consists of 
a representation $\bar 5\oplus 10$, then the absence of anomalies necessitates 
a number of generations multiple of 3, with proper chiralities, see 
Eq.\ (\ref{SU5_gen}). 
However, note that anomalies do not impose that a generation be 
$\bar 5 \oplus 10$; one could as well choose $5 \oplus 10$, or any such 
combination, although the condition remains $n_g=0$ mod 3. 
In fact, we have mentionned that the minimal anomaly-free fermion content 
for $SU(5)$ in $D=6$ is three 5 or $\bar 5$ with identical chirality, plus a 
10 or $\overline{10}$ with opposite chirality. 
This means that anomalies are not the ultimate answer; there must be another 
ingredient which must be combined with anomalies. 

This additional ingredient might be some principle requiring that the theory 
be built with identical (including the chirality assignment) building blocks, 
as the Standard Model in four dimensions. 
In that case, the only anomalies which can give some restriction on the number 
of such building blocks are the global anomalies, both gauge and, to a lesser 
extent, gravitational. 
Stated differently, global anomalies are the only ones which can impose to 
have identical copies of a basic, necessarily locally anomaly-free, 
generation. 
Under this assumption, the only theories with anomaly-free fermion contents 
are $\SM$ in $D=6$, with $n_g=0$ mod 3 \cite{Dobrescu:2001ae}, and 
six-dimensional $E_6$, eight- and ten-dimensional $\SM$, and $SU(5)$ in 
$D=10$, all of which are anomaly-free whatever the number of identical 
generations. 

Our six-dimensional $SU(5)$ solution, where the number of generations is 
imposed by the absence of the local gauge anomaly, does not satisfy this
criterion.  
Nonetheless, the nonidentity of the generations might be a blessing and could 
give rise to interesting phenomenological predictions, for instance regarding 
the hierarchy of fermion masses in the Standard Model.

\section*{Acknowledgements}

We wish to thank Jean-Marie Fr\`ere for useful discussions and suggestions. 
N.\ B.\ and Y.\ G.\ acknowledge support from the ``Actions de Recherche 
Concert\'ees'' of the ``Communaut\'e Fran\c{c}aise de Belgique'' and 
IISN-Belgium.

\appendix

\section{Chirality in even dimensions}
\label{s:C,P,etal}

To fix the notations we use throughout the paper, and for completeness sake, 
we recall in this Appendix some basic definitions and properties about chiral 
fermions in Euclidean and Minkowski space \cite{Alvarez-Gaume:1985ex}. 

\subsection{Chirality and even dimensions}
\label{s:chirality}

The Lorentz group will be denoted $SO(t,s)$ where $t$ and $s$ are the numbers 
of time and space dimensions respectively $(D=t+s)$. 
 A Dirac spinor obeys the following transformation law under infinitesimal 
Lorentz transformations :
\begin{equation}
  \label{eq:spinorlorenz}
  \delta \Psi \equiv - \frac{1}{2} \omega_{MN} \Sigma_{(D)}^{MN} \Psi,  \qquad 
M,N=0,\ldots, D-1, 
\end{equation}
where  $\omega_{MN}$ are real coefficients and $\Sigma_{(D)}^{MN} $ denote  
the Lorentz generators in spinorial representation. 
The latter are  given in terms of Dirac $2^{[D/2]} \times 2^{[D/2]}$ matrices
$\Gamma_{(D)}$, where $[A]$ is the integer part of $A$, which satisfy the 
Clifford algebra:
\begin{eqnarray}
  & \Sigma_{(D)}^{MN} \equiv \frac{1}{4} \left[ \Gamma^M_{(D)} , 
    \Gamma_{(D)}^N  \right], &  \label{Sigma^MN} \\
  & \left\{ \Gamma_{(D)}^M , \Gamma_{(D)}^N  \right\} = 2 \eta_{MN}. & 
  \label{Clifford}
\end{eqnarray}
The matrix $\eta_{MN}$ is the flat metric with signature $(t,s)$. 
In the following, we shall drop the subscript $(D)$, except when there may be  
some ambiguity. 

In any even dimension $D$, one can introduce the matrix $\bar\Gamma$, which 
is the analog of $\gamma_5$ in $D=4$:
\begin{equation}
  \bar \Gamma \equiv \alpha \Gamma^0 \Gamma^1 \cdots \Gamma^{D-1}
\end{equation}
where the coefficient $\alpha$ is conventionally chosen such as 
$(\bar \Gamma)^2=1$. 
This matrix anticommutes with all Dirac matrices, and thus commutes with all 
Lorentz generators $\Sigma^{MN} $. 
The latter property means that the Dirac representation is reducible: a Dirac 
spinor $\Psi$ splits in two irreducible parts $\Psi_+$ and $\Psi_-$, called 
Weyl spinors, which transform independently under Lorentz transformations.
The Weyl spinors are defined as follows:
\begin{equation}
  \label{eq:weylspinor}
  \Psi_{\pm} \equiv \frac{1}{2} (1 \pm \bar \Gamma ) \Psi. 
\end{equation}
Throughout the paper, we replace $\Psi_+$ and $\Psi_-$ with $\Psi_{R_D}$ and 
$\Psi_{L_D}$ respectively. 

In odd dimension $D$, the set of Dirac matrices consists of the $D-1$ Dirac 
matrices in dimension $D-1$, plus an additional one which is proportionnal to 
$\bar\Gamma_{(D-1)}$.  
Thus, it is no longer possible to have another matrix which could anticommute 
with all matrices $\Gamma_{(D)}$. 
This prevents the definition of chirality (and therefore chiral fermions or 
anomalies) in odd dimensions.

\subsection{Charge conjugation in $4k$ and $4k+2$ dimensions}
\label{s:C}

Since there is only one faithful representation of the Clifford algebra with a 
given dimension,  all sets of matrices which satisfy Eq.\ (\ref{Clifford}) 
are related by similarity transformations. 
As the set of the complex conjugate matrices $(\Gamma^M)^{*}$ fulfils this 
requirement, there exists a matrix $B$ such that:
\begin{equation}
  \label{eq:matrixB}
  \left(\Gamma^M\right)^*=B  \Gamma^M B^{-1}, \qquad
  \left(\Sigma^{MN}\right)^*=B  \Sigma^{MN} B^{-1}
\end{equation}
Using this relation, one can define a charge conjugate Dirac spinor $\Psi^c$ 
which transforms exactly in the same way as $\Psi$  
[Eq.\ (\ref{eq:spinorlorenz})]:
\begin{eqnarray}
  \label{eq:psiC}
  \Psi^c & \equiv & {\sf C} \Psi \equiv B^{-1} \Psi^{*} \\
  \delta (\Psi^c) & \equiv & - \frac{1}{2} \omega_{MN} \Sigma^{MN} \Psi^{c}
\end{eqnarray}
where {\sf C} is the charge conjugation operator. 

The transformation of a chiral, Weyl fermion under charge conjugation depends 
on the numbers of space and time dimensions. 
When $(s-t)/2$ is odd, $\{ {\sf C} , \bar\Gamma \}=0$, so that charge 
conjugation flips chirality; otherwise, {\sf C} and $\bar\Gamma$ commute, 
and charge conjugation does not modify  the chirality of a Weyl fermion. 
In the ``usual'' case with a single time dimension $t=1$, which we assume from 
now on, this gives: 
\begin{eqnarray}
  \label{eq:4k4kplus2}
& \{ {\sf C} , \bar\Gamma \}=0 & \mbox{ in } D=4k, \cr 
& [{\sf C} , \bar\Gamma ]=0 & \mbox{ in } D=4k+2.
\end{eqnarray}
{\sf C} flips chirality in $D=4k$, while it does not in $D=4k+2$.

\subsection{Chiral representation}
\label{s:gamma}

We recall a possible explicit realization of the Dirac $\Gamma$ matrices for 
a spacetime dimension $D$, which has the attractive feature to separate left- 
and right-handed fermions. 
This representation will also be useful for the trace calculation in 
Appendix \ref{s:Tr(SO10)^5}.

In dimension $D$, the matrices are built from the Dirac matrices in dimension 
$D-1$ following:
\begin{eqnarray}
  \label{gammachiral}
\Gamma^0_{(D)}\! &=&\! 
\left[ \begin{array}{cc} 0 & {\bf 1}_{D/2} \\
{\bf 1}_{D/2} & 0 \end{array} \right] \, ,\cr
\Gamma^k_{(D)}\! &=& \! 
\left[ \begin{array}{cc} 0 & \Gamma^0_{(D-1)} \Gamma^k_{(D-1)} \\ 
-\Gamma^0_{(D-1)} \Gamma^k_{(D-1)} & 0 \end{array} \right] \, ,\cr
\Gamma^{D-1}_{(D)} \! &=& \! 
\left[ \begin{array}{cc} 0 & \Gamma^0_{(D-1)}  \\ 
-\Gamma^0_{(D-1)}  & 0 \end{array} \right] \, ,{\rm and} \cr
\bar \Gamma_{(D)}\! &=& \! 
\left[ \begin{array}{cc} {\bf 1}_{D/2} & 0   \\ 
0 & -{\bf 1}_{D/2} \end{array} \right] \, ,
\end{eqnarray}
where $k = 1,\dots, D-2$ and ${\bf 1}_{D/2}$  is the $D/2$-dimensional unity 
matrix.
As recalled above, the matrices $\Gamma^k_{(D-1)}$ where $k$ runs from 1 to 
$D-3$ are the Dirac matrices $\Gamma^k_{(D-2)}$ in dimension $D-2$, and 
$\Gamma^{D-2}_{(D-1)}=i \bar\Gamma_{D-2} $.
Since the Lorentz generators $\Sigma_{(D)}^{MN}$, Eq.\ (\ref{Sigma^MN}), are 
commutators of Dirac matrices, they are all block diagonal, made of two 
$D/2 \times D/2$ blocks. 
These blocks yield the generators of the transformation for the $\Psi_{L_D}$ 
and $\Psi_{R_D}$.

\section{$SO(10)$ traces in 8 dimensions}
\label{s:Tr(SO10)^5}

In this Appendix, we show that the symmetrized trace over the product of 
five $SO(10)$ generators does not vanish in general, and more precisely is 
proportional to the Levi-Civita tensor with 10 indices. 
This is analogous to the fact that in $D=4$ dimensions, the triangle anomaly 
vanishes for every group $SO(N)$ except if $N=6$, the trace over three 
$SO(6)$ generators being proportional to the six-indices Levi-Civita tensor.

We are interested in symmetrized traces which involve 5 generators $T$ for 
the basic $10$ and the spinorial $16$ representations of $SO(10)$, which yield 
the local gauge anomalies of the corresponding representations in $D=8$. 
These symmetrized traces, Eq.\ (\ref{gauge8D}), can be written
\begin{equation}
\label{TrSO10}
{\rm STr} \left( (T^a)^{\alpha\beta} (T^b)^{\gamma\delta} (T^c)^{\kappa\lambda}
(T^d)^{\mu\nu} (T^e)^{\rho\sigma}\right), 
\end{equation}
where all Greek indices run from 0 to 9. 
The symmetrized trace is invariant under orthogonal $SO(10)$ transformations 
of the generators. 
In terms of Minkowski instead of Euclidean spacetime, the trace 
Eq.\ (\ref{TrSO10}) is a $SO(1,9)$ invariant, which means that it is 
proportional to a 10-indices tensor of a $D=10$ theory with appropriate 
symmetry properties.  
The only possible one is the totally antisymmetric Levi-Civita tensor 
$\epsilon_{\alpha\beta\gamma\delta\kappa\lambda\mu\nu\rho\sigma}$. 
We only have to check whether the proportionality constant is zero or not, 
for each representation we consider. 

The representation $10$ is real [by definition of $SO(10)$], and thus all 
generators are similar to their complex conjugate: $ S T S^{-1}=-T^*$. 
Since the trace involves the product of 5 generators, it is equal to the 
opposite of the trace over the conjugate generators. 
On the other hand, the generators are hermitian, so that both traces over 
either the generators or their complex conjugates are equal. 
Therefore, the trace Eq.\ (\ref{TrSO10}) vanishes for the 10. 

The (Dirac) spinor, reducible 32 representation of $SO(10)$ is also a 
representation of $SO(1,9)$, generated by the $\Sigma^{MN}_{(10)}$, and we may 
use the choice of generators of Sec.\ \ref{s:gamma}, in the case $D=10$. 
These are antisymmetric, $32 \times 32$ matrices, composed of two 
$16 \times 16$ blocks, which transform $D=10$ left- and right-handed Weyl 
fermions, which are precisely the 16 and $\overline{16}$ of $SO(10)$. 
As noted in previous section, these blocks involve the $D=8$ Dirac matrices 
$\Gamma^M_{(8)}$, plus $\bar\Gamma_{(8)}$.  
Let us take the upper block to describe our $16$ representation. 
Calculating the trace of a product of $16$ is now straightforward: we just 
compute the product of the $32 \times 32$ matrices $\Sigma^{MN}_{(10)}$ and 
then evaluate the trace over the upper block of the product. 

Consider the product 
$\Sigma^{01}_{(10)}\Sigma^{23}_{(10)}\Sigma^{45}_{(10)}\Sigma^{67}_{(10)}
\Sigma^{89}_{(10)}$.
The upper block is proportional to :
\begin{equation}
\Gamma^0_{(8)} \Gamma^1_{(8)} \cdots \Gamma^7 _{(8)}\bar\Gamma_{(8)} 
\propto \left( \bar \Gamma_{(8)}\right)^2,
\end{equation}
which is the identity matrix ${\bf 1}_{16}$. 
Thus, the trace is non-zero, and so is the proportionality coefficient with 
the Levi-Civita tensor.

\end{document}